\documentclass[letter]{aa} 
\usepackage{graphicx}
\usepackage{txfonts}
\usepackage{natbib}
\def\hhhop{H$_3$O$^+$}
\def\hho{H$_2$O}

\def\hh{H$_2$}

\def\nnhp{N$_2$H$^+$}

\def\cop{CO$^+$}

\def\meth{CH$_3$OH}
\def\msol{M$_{\odot}$}

\def\pow#1#2{#1$\times$10$^{#2}$}
\def\smm{(sub-)mil\-li\-me\-ter}

\def\gtsim{{_>\atop{^\sim}}}
\def\ltsim{{_<\atop{^\sim}}}

\def\kms{km~s$^{-1}$}
\def\tas{$T_A^*$}

\def\txc{$T_{\rm ex}$}

\def\vhel{V$_{\rm hel}$}
\def\ccm{cm$^{-3}$}
\def\scm{cm$^{-2}$}

\def\rs{s$^{-1}$}
\def\dv{$\Delta$\textit{V}}
\def\mic{$\mu$m}
\begin{document}
\title{Detection of extragalactic \hhhop}

\author{F.F.S. van der Tak\inst{1,2} \and S. Aalto\inst{3} \and
  R. Meijerink\inst{4}}

\institute{SRON Netherlands Institute for Space Research,
           Landleven 12, 9747 AD Groningen, The Netherlands;
           \email{vdtak@sron.nl} \and
          Kapteyn Astronomical Institute, University of Groningen, The Netherlands
          \and Onsala Space Observatory, Chalmers University of Technology, 43992
          Onsala, Sweden \and
          Astronomy Department, University of California, 601 Campbell Hall,
          Berkeley, CA 94720, USA}

\date{Received 10 October 2007 ; accepted 12 November 2007}

\titlerunning{Detection of extragalactic \hhhop}
\authorrunning{F.F.S. van der Tak et al.}

\abstract
{The \hhhop\ molecule probes the oxygen chemistry and the
  ionization rate of dense circumnuclear gas in galaxies.}
{Recent \hhhop\ observations show variations in the
  cosmic-ray ionization rate by factors of $>$10 within our Galaxy.}
{Using the JCMT, we have observed the 364 GHz line of p-\hhhop\ in the centers of
  M82 and Arp 220.}
{In Arp 220, the line profile suggests that the emission originates in the Western
  nucleus. In M~82, both the eastern molecular peak and the circumnuclear region
  contribute to the emission. The derived column densities,
  abundances, and \hhhop\ / \hho\ ratios indicate ionization rates similar to
  or even exceeding that in the Galactic Center.}
{ Model calculations of the chemistry of irradiated molecular gas indicate a
  likely origin of this high ionization rate in the extended, evolved starburst
  of M82. In contrast, irradiation by X-rays from the AGN disk is the most
  likely model for Arp~220.}

\keywords{galaxies: starburst; galaxies: active; radio lines: galaxies; ISM: molecules} 

\maketitle
%

\section{Introduction}
\label{s:intro}

Starbursts and active galaxies often host large masses of molecular gas and dust
in their inner kpc. Gravitational interaction or merging cause the funneling
(via bars) of the gas towards the center. The average gas surface- and number
densities are significantly higher (by several orders of magnitude) than in the
more extended disks. The nuclear molecular gas is often found in tori, feeding
the activity and thus controlling its evolution. The triggering and turn-off
mechanisms of the activity are dependent on the properties of the gas, which
are still poorly known, in particular the fraction of the gas at high
temperature and density.  Observations of the molecular gas properties are
starting to show great promise not only in allowing us to model the evolution of
the starburst, but also in helping to identify and analyze the type of
activity lurking in very deeply obscured galactic nuclei. Here, dust extinction
is high enough to quench emission even down to the mid-infrared.
For reviews of this subject, see \citet{sanders:ulirgs} and \citet{gao:ulirgs}.

The star formation rates in the nuclei of nearby galaxies are measured best at 
\smm\ wavelengths, where extinction is negligible and where a rich spectrum of dust
continuum and molecular lines allows us to measure fundamental parameters of
star-forming matter such as mass, temperature, and density (e.g., \citealt{aalto:ulirgs}).
However, existing observations are mostly at millimeter wavelengths and thus
limited to relatively low temperatures and densities.
The excellent transmission of sites like Mauna Kea and Chajnantor enables
observation of the \hhhop\ molecule, which has a twofold importance.

First, the \hho\ molecule acts as a natural filter for selecting
hot ($\ga$100\,K) gas, and therefore traces different regions than CO and
other molecules commonly studied from low-altitude sites. The chemistry of this
filtering is the evaporation of icy grain mantles at $T\approx 100$\,K
\citep{vdtak:h2o}.
Other molecules evaporating from grain mantles are not as abundant as \hho.
The \hhhop\ molecule is an excellent proxy for \hho, which cannot be
observed in thermal lines from the ground at redshifts below 0.01. 
In contrast to HDO whose ratio to \hho\ strongly depends on ambient conditions,
the chemical relation
between \hhhop\ and \hho\ is well-understood and the relevant reactions have
been measured in the laboratory \citep{phillips:h3o+}. Thus, observations of
\hhhop\ are useful probes of the gas distribution and kinematics in
the warm and dense nuclei of active galaxies. 

Second, the high proton affinity of \hho\ makes \hhhop\ a key ion in the oxygen
chemistry of dense molecular gas. Combined with information on the \hho\
abundance, observations of \hhhop\ may be used to trace the ionization rate
of dense circumnuclear clouds by cosmic rays (produced in supernovae, i.e.,
starbursts) and X-rays (from an AGN). The ionization rate of this gas is a
fundamental parameter for its dynamics, because it regulates the efficiency of
magnetic support against gravitational collapse. 
Recent APEX observations of \hhhop\ show that the cosmic-ray ionization
rate in the Sgr~B2 cloud near the Galactic center is $\sim$10$\times$ higher
than in star-forming regions at 1--4\,kpc from the Sun \citep{vdtak:sgrb2}.

Motivated by the results of \hhhop\ observations in our Galaxy, we searched
for \hhhop\ emission in two prototypical active galaxies: M82 and Arp~220. 
At a distance of 4\,Mpc, M82 is the prototype evolved starburst, with a large
\hh\ mass, where many molecules were seen for the first time outside our
Galaxy, including \cop, a tracer of strong irradiation of molecular gas
\citep{fuente:co+}.
The molecular emission can be roughly divided into two prominent lobes of
molecular emission, separated by 30$''$, which is interpreted as a 400 pc
rotating molecular ring. There is also molecular gas in the nucleus, which is
best seen in high-$J$ CO lines or in high-density tracer emission, such
as HCO (e.g., \citealt{garcia-burillo:m82}).

The prototypical ultraluminous galaxy Arp~220 ($d$ = 72\,Mpc), where water lines
have been detected \citep{gonzalez:arp220,cernicharo:arp220}, is a merger of two
gas-rich disk galaxies. Significant ($\gtsim$10$^9$\,\msol) molecular gas has
gathered in two nuclei (east and west) that have not yet merged into one. The
two nuclei are separated by 1$''$ and are surrounded by a 5$''$ extended disk of
molecular gas. The bulk of the molecular emission from Arp~220 is emerging from
this region (e.g., \citealt{sakamoto:arp220}), and the JCMT beam covers all of
it.
Arp 220 has a rich chemistry with surprisingly bright HNC 3--2 line emission,
outshining HCN 3--2 by a factor of 2, which indicates unusual chemical or
excitational circumstances. High-resolution observations reveal that the HNC
emission is emerging from warm gas ($T_{\rm k} >$ 40\,K) near the western nucleus
\citep{aalto:madrid}.


\section{Observations}
\label{s:obs}

The $3_2^+$--$2_2^-$ line of \hhhop\ was observed during April 26-28, 2007, with
the James Clerk Maxwell Telescope (JCMT)\footnote{The JCMT is operated by the
  Joint Astronomy Center on behalf of the Science and Technology Facilities
  Council of the United Kingdom, the Netherlands Organization for Scientific
  Research, and the National Research Council of Canada.} under program M07AN03.
In the 345\,GHz window, the telescope has a beam size of 14$''$ and a main beam
efficiency of 70\%, as determined by the telescope staff and verified by scans of
Mars at the end of each night.
One of the central pixels of the Heterodyne Array Receiver Program (HARP) was
used as front end, with receiver temperatures of $\sim$120\,K at our observing
frequency of $\sim$360\,GHz.
The back end was the Auto-Correlation Spectrometer and Imaging System (ACSIS),
providing 1.0\,GHz bandwidth in 2048 channels.
The weather was excellent throughout, with water vapour columns of
$\ltsim$1\,mm, corresponding to a zenith opacity of $\tau \ltsim 0.05$ at
225\,GHz. System temperatures were 300-400\,K for M82 and 200-300\,K for
Arp~220, which reaches higher elevations and whose larger redshift puts the line
at a frequency with better atmospheric transmission.
Double beam switching at a rate of 1\,Hz was used with an offset of 120$''$ in
azimuth, which led to flat, stable baselines.
Integration times (on+off) are 8.2\,hrs for Arp~220 and 6.5\,hrs for M82,
resulting in rms noise levels of \tas\ = 2-3\,mK per 8\,\kms\ channel. 
Telescope pointing was checked every hour on the CO line emission of nearby AGB
stars and always found to be within 2$''$. The telescope z-focus was checked on
bright AGB stars at the beginning of the night (after sunset) and around
midnight, and corrected by 0.1-0.2\,mm. The bright AGB stars were also used to
take standard spectra, and the results lead us to believe that the calibration
of the data is accurate to $\sim$10\%.

\begin{figure}[tb]
\centering
\includegraphics[width=5cm,angle=-90]{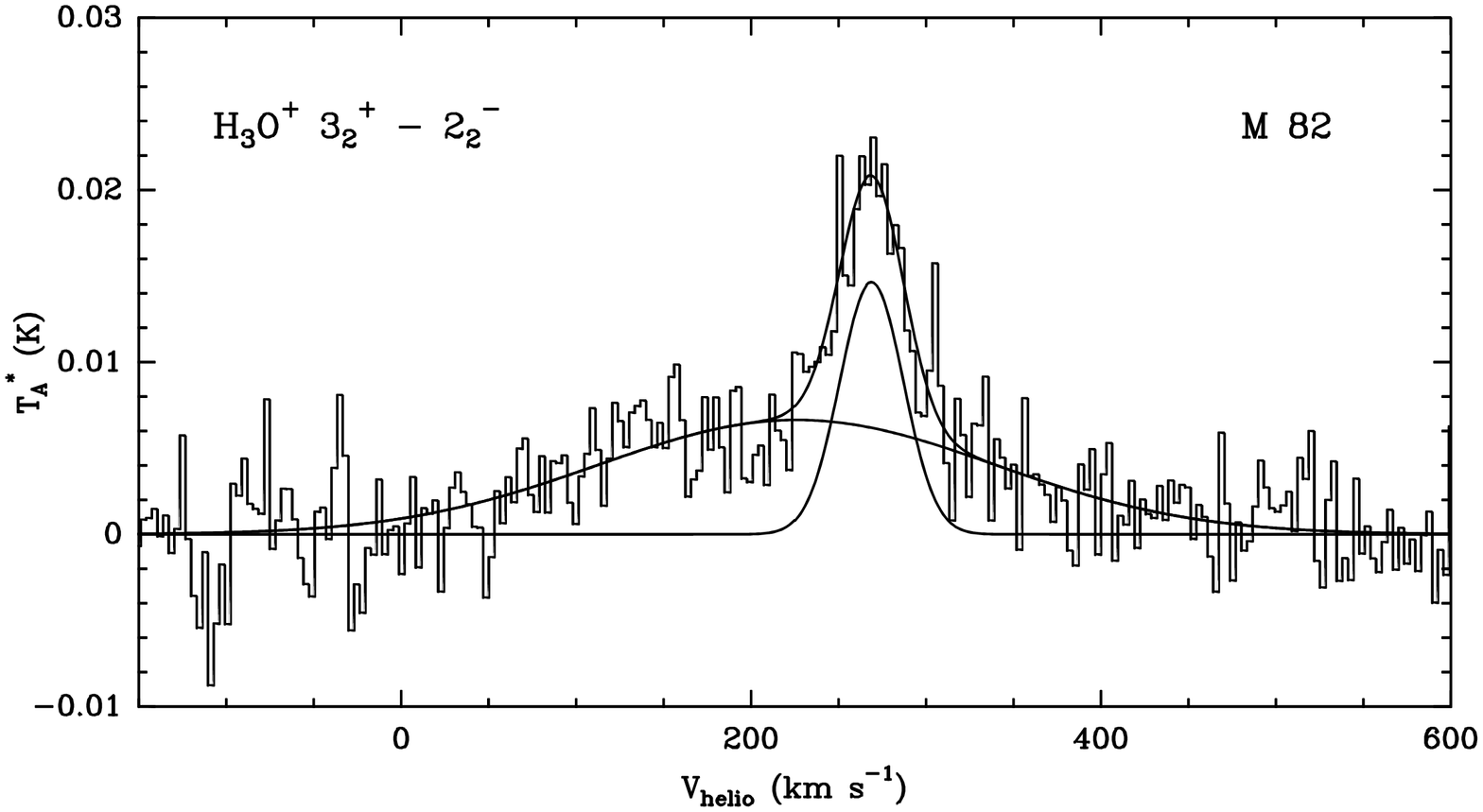}
\bigskip
\includegraphics[width=5cm,angle=-90]{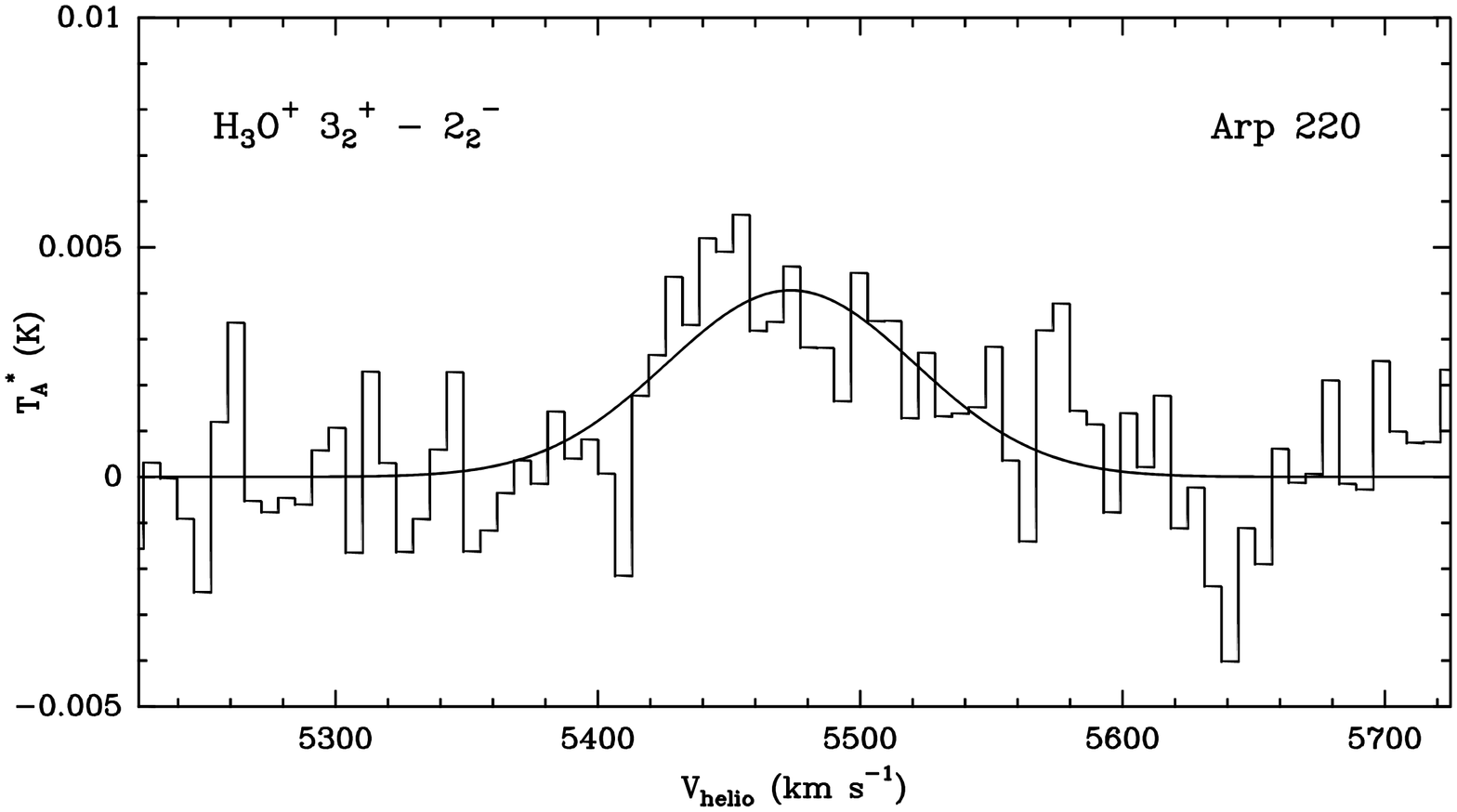}  
\caption{Spectra of the \hhhop\ $3_2^+$--$2_2^-$ line toward M82 (top) and
  Arp~220 (bottom), observed with the JCMT.} 
\label{f:spec}
\end{figure}

Initial reduction of the data was done with the Starlink package, and the final
analysis performed in Class. We inspected the individual scans, summing the good
ones, and subtracted linear baselines.
The identification of the spectral feature as \hhhop\ is secure, since the
standard molecular line catalogs (\citealt{pickett:jpl}; \citealt{mueller:cdms})
do not list any other plausible candidates at this frequency, and the
attenuation of the image sideband is better than 10\,dB.
After smoothing the data to a velocity
resolution of 8-16\,\kms, line parameters were extracted by fitting Gaussian
profiles to the spectra. The final spectra are shown in Fig.~\ref{f:spec} and
the line parameters listed in Table~\ref{t:line}.
Throughout this paper, velocities are in the heliocentric frame and redshift is
computed using the radio convention.

\begin{table}[tb]
\caption{Measured line parameters with 1$\sigma$ errors in brackets in units of
  the last decimal.} 
\label{t:line}
\begin{tabular}{lllrrr}
\hline
\hline
\noalign{\smallskip}
Source & $\alpha$(J2000) & $\delta$(J2000) & \tas    & \vhel   & \dv \\
          & hh mm ss     & $^0$ $'$ $''$   & mK      & \kms    & \kms \\
\noalign{\smallskip}
\hline
\noalign{\smallskip}
M82       & 09:55:52.19  & $+$69:40:48.8   & 15(2)  & 269(2)  &  43(5) \\  
 & \multicolumn{2}{l}{broad component:} & 6.5(6) & 224(11) & 261(23) \\
Arp 220   & 15:34:57.23  & $+$23:30:11.5   & 4.1(6) & 5473(8) & 113(20) \\  
\noalign{\smallskip}
\hline
\end{tabular}
\end{table}

\section{Results}
\label{s:results}

Emission due to \hhhop\ is clearly detected towards the nuclei of both M82 and
Arp~220.  The signal-to-noise ratio on the line profiles is high enough to
address the location of the emission within the galaxies. 
In the case of Arp~220, the line profile is single-peaked and similar to that of
\meth\ \citep{jesus:paris}. The emission peak at
\vhel$\approx$5500\,\kms\ indicates an origin in the western nucleus, which is
also where the HNC 3--2 emission peaks in recent interferometric images \citep{aalto:hnc}.
In the case of M82, the emission peak at \vhel$\approx$300\,\kms\ suggests an
origin in the eastern molecular peak at about 12$''$ (240~pc) from the nucleus,
which is where the CO emission peaks \citep{tilanus:m82} and where \cop\ and
\meth\ are also located \citep{fuente:co+,martin:meth}. The profile almost
extends to \vhel$\approx$200\,\kms, which suggests that significant \hhhop\ is
also present in the central region of the galaxy, as found before for \nnhp\
\citep{mauersberger:n2h+}.
In fact, a single Gaussian profile does not give a good fit to the M82 spectrum.
Figure~\ref{f:spec} shows a decomposition of the profile into two Gaussians, where
the narrow component represents \hhhop\ in the eastern molecular peak of M82 and
the broad component either the more widely-distributed molecular gas or the
contribution from the nucleus. This decomposition of the line profile may not be
unique, but serves to illustrate the existence of multiple \hhhop\ components in
the nucleus of M82.
Although the JCMT beam is too small to cover the eastern lobe of M82 while
pointing at the center, it is possible that we are picking up emission from the
inner side of the lobe. The other HARP pixels do not cover the lobe either, but
future fully-sampled maps will reveal the spatial distribution of \hhhop\ in
M82.

To estimate the column density of \hhhop\ towards our sources from the
integrated line intensities (Table~\ref{t:line}), we adopt a main
beam efficiency of 70\% and regard the expected kinetic temperature of
  $\sim$100\,K as an upper limit to the excitation temperature (\txc).
For \txc\ = 50 -- 100\,K, the column density of p-\hhhop\ is \pow{0.7 --
1.3}{13}\,\scm\ for Arp~220 and \pow{3.0 -- 5.5}{13}\,\scm\ for M82. The
total beam-averaged column density of \hhhop\ is 2 -- 3 times higher, because
the ortho to para ratio of \hhhop\ drops from the high-temperature limit of
$o/p = 1$ at $T \gtsim 100$\,K to $o/p \gtsim 2$ at $T \ltsim 50$\,K. The
following discussion adopts the intermediate case of $o/p = 1.5$; we refer to
\citet{phillips:h3o+} and \citet{vdtak:sgrb2} for further discussion of
\hhhop\ radiative transfer.

Unlike for lower-frequency lines of heavier molecules that are excited by
collisions, the excitation temperature of \hhhop\ probably reflects the colour
temperature ($T_C$) of the far-infrared radiation field rather than the kinetic temperature
of the molecular gas. The low reduced mass of the \hhhop\ molecule makes its
excitation very sensitive to radiative pumping by dust \citep{phillips:h3o+,vdtak:sgrb2}.
Calculations with the \texttt{RADEX} non-LTE radiative transfer program
\citep{vdtak:radex}
using molecular data from the LAMDA database \citep{schoeier:lamda}
indicate that the coupling between
\txc\ and $T_C$ holds as long as $T_C > 30$\,K and that the measured line is
optically thin under these conditions. We assume a filling factor of the dust
radiation field of unity, representing the case that the gas and dust are well-mixed.


The beam-averaged \hhhop\ column density in Arp~220 is almost certainly an
underestimate of its source-averaged value,
because the emission is unlikely to be as extended as 14$''$.
First, the molecular gas traced by the CO 2--1 emission is concentrated in two
nuclei of diameter 300~pc each \citep{sakamoto:arp220}. Recent high-resolution
work on HNC 3--2 \citep{aalto:hnc} further shows that the dense gas is
concentrated towards the western nucleus with a source size of 0.7~arcsec
(260~pc).
Second, since \hhhop\ is excited by infrared radiation, its emission may follow
the infrared light distribution. From 3--25~\mic\ Keck observations,
\citet{soifer:arp220} find a size of 0.73$''$ at 25~\mic. They derive a size of
2$''$ at 100~\mic, which seems a firm upper limit to the \hhhop\ source
size. 
Furthermore, the absence of a double-peaked profile from the two nuclei in our
data suggests that the emission is located on one of the two nuclei and does
not cover the full extent of the CO emission. We thus adopt a source
size of 0.7$''$, which must of course be verified at high resolution.

In the case of M82, the dense molecular gas is distributed over many clumps as shown
by high-resolution CO and HCO images \citep{garcia-burillo:m82}.  These clumps
are grouped in three lobes (east, middle, and west) separated by $\approx$15$''$
(300~pc), with significant emission coming from the medium in between.  Our JCMT
beam is pointed to a position close to the central lobe, so that a significant
fraction of the beam is probably filled with \hhhop\ emission.


Table~\ref{t:coldens} reports the column densities of \hhhop, \hho, and \hh\ in
our sources, as taken from this work and the literature. 
The values refer to a $\approx$15$''$ area for M82 and a $\approx$0.7$''$
  area for Arp 220.
Our adopted $N$(\hh) toward Arp~220 is 10$\times$ below the peak value reported
by \citet{downes:arp220} for a 0.2$''$ area, so that it is representative of
our assumed \hhhop\ source size of 0.7$''$.
The table shows that the abundance of \hhhop\ relative to \hh\ is
$\sim$\pow{2--10}{-9} in these sources.
These values are much higher than the typical value for Galactic star-forming
cores \citep{phillips:h3o+} and at or above the value of \pow{3}{-9} in the Sgr
B2 cloud at the Galactic center \citep{vdtak:sgrb2}. The ionization rate of the
molecular gas in the nuclei of Arp~220 and M82 thus seems comparable to the
value in the Galactic center.
This conclusion is supported by the \hhhop/\hho\ ratio of 1/50 that we find for
Arp~220, which is similar to the value of 1/20 for the Galactic center. No useful
estimate of this ratio can be made for M82: the Odin data quoted in
Table~\ref{t:coldens} have a beam size of 2$'$ and thus sample a very different
gas volume than our JCMT data.
Note that our estimated \hhhop\ abundance in Arp~220 is somewhat hampered by
source size uncertainties.

\begin{table}[tb]
\caption{Source-averaged column densities [\scm] of \hh, \hho, and \hhhop.} 
\label{t:coldens}
\begin{tabular}{lrrrc}
\hline
\hline
\noalign{\smallskip}
Source    & $N$(\hh)    & $N$(\hho)      &  $N$(\hhhop)  &  Refs. \\
\noalign{\smallskip}
\hline
\noalign{\smallskip}
M82       & \pow{1}{23} &$<$\pow{2}{14}  & \pow{1.1}{14} & [1,2,3] \\
Arp 220   & \pow{1}{24} &   \pow{2}{17}  & \pow{1.0}{16} & [4,5,6] \\
\noalign{\smallskip}
\hline
\noalign{\smallskip}
\multicolumn{5}{r}{References: [1] \cite{mao:m82}; [2] \cite{wilson:odin};}\\
\multicolumn{5}{r}{[3] This work; [4] \cite{downes:arp220};} \\
\multicolumn{5}{r}{[5] \cite{gonzalez:arp220}; [6] This work.}
\end{tabular}
\end{table}


\section{Discussion}
\label{s:disc}

\begin{figure*}[tb]
\centering
\includegraphics[width=5cm,angle=0]{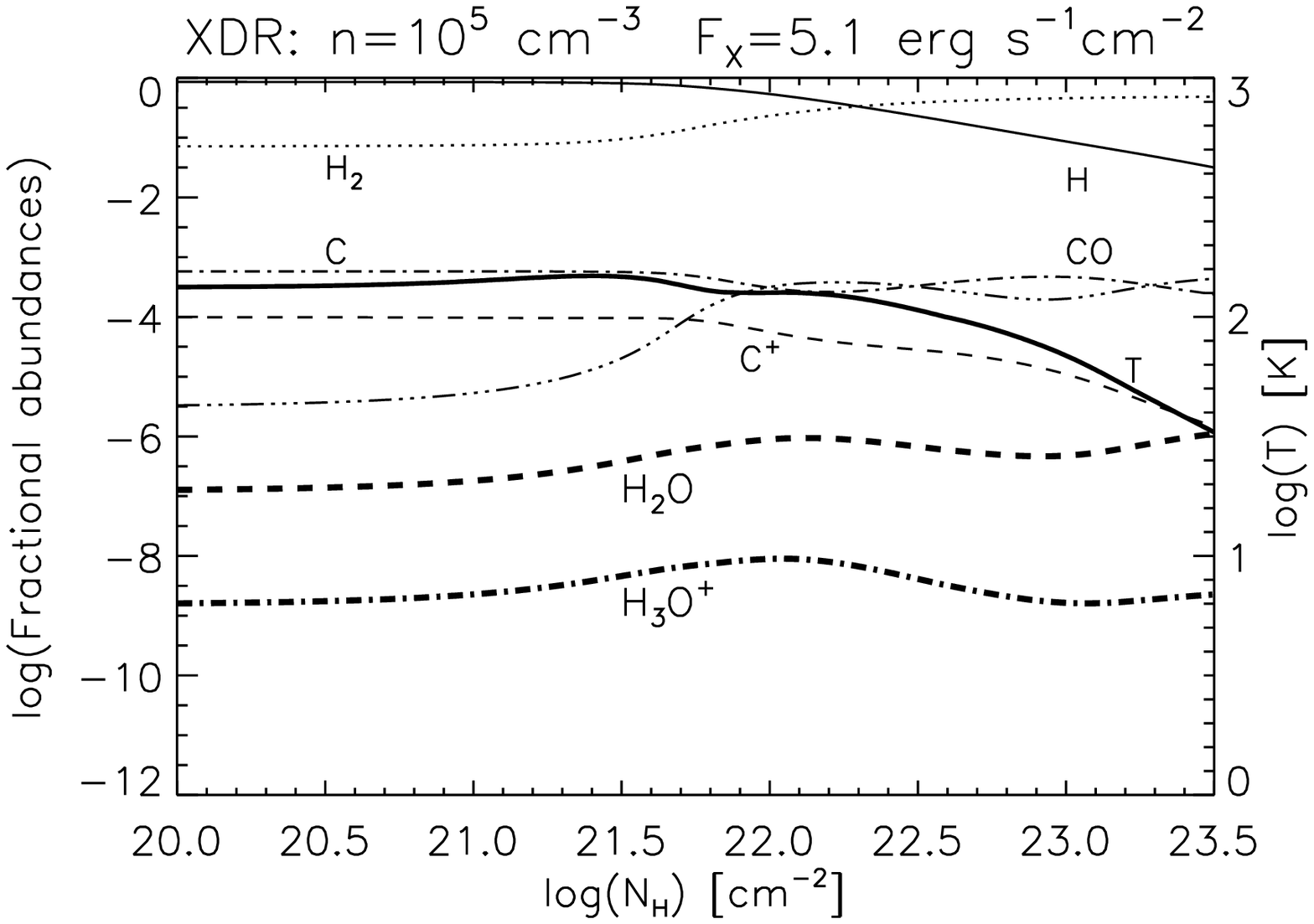}
\bigskip
\includegraphics[width=5cm,angle=0]{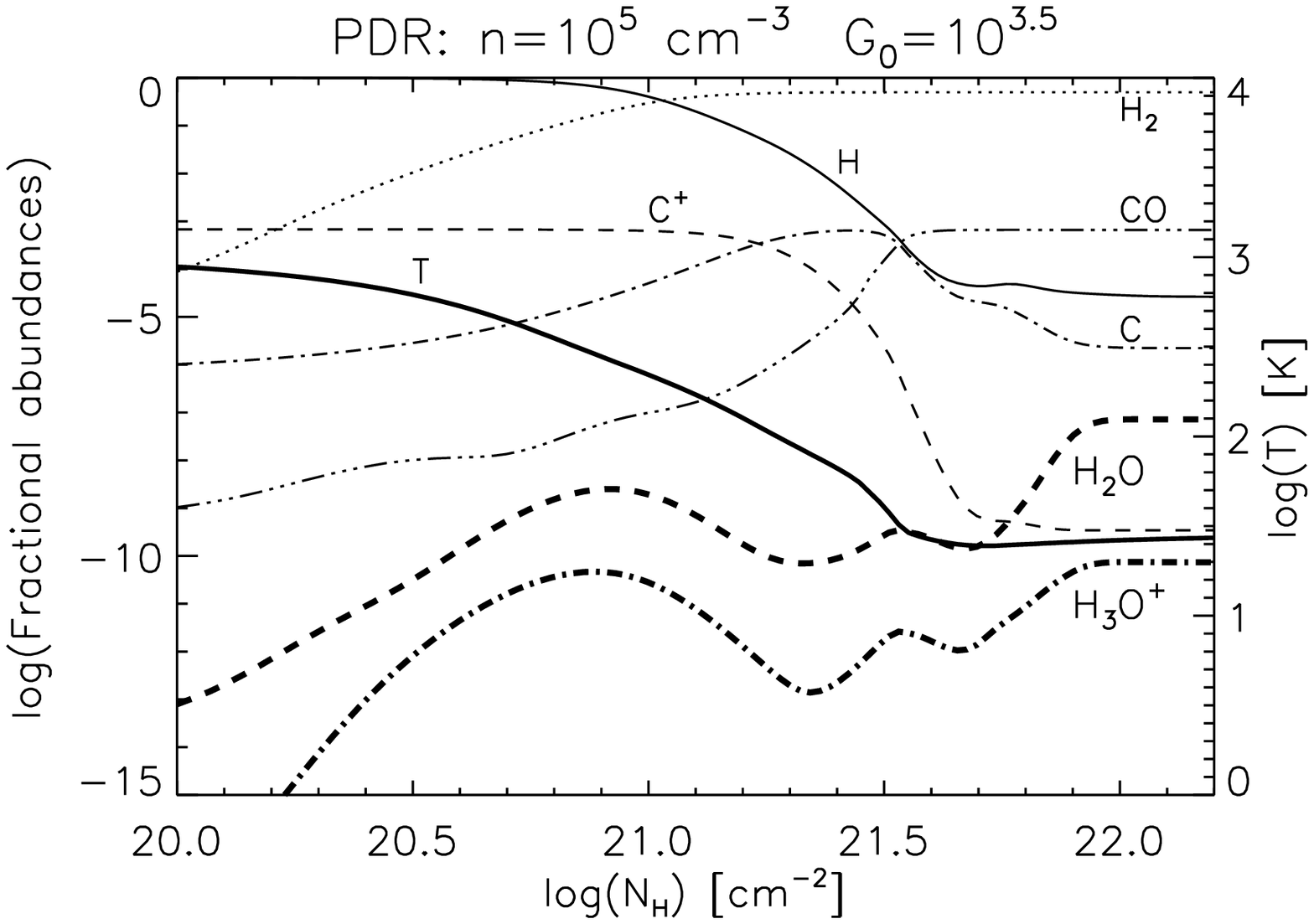}  
\caption{Calculated chemical structure of XDR (top) and PDR (bottom) models}.
\label{f:model}
\end{figure*}

The column densities of \hhhop, \hho, and \hh\ in Table~\ref{t:coldens} and
their ratios are compared to chemical models of clouds irradiated by either
far-UV or X-ray photons. Far-UV dominated clouds are called photon-dominated
regions (PDRs) and the other type X-ray dominated regions (XDRs). The thermal
and chemical structures of PDRs and XDRs are quite different, because the
absorption cross section for X-rays is much lower than for far-UV. The
ionization fraction in XDRs can be as high as $x_e = 10^{-2} - 10^{-1}$, while
in PDRs, it is 10$^{-4}$ at most. Furthermore, in XDRs, larger parts of the
cloud can be kept at a high temperature.
For an elaborate discussion of the differences between
PDRs and XDRs we refer to \citet{meijerink:program} where the codes are
described. The grid of models by \citep{meijerink:grid} is used to look for
the best match to our observations. For the PDR models, we also consider
an elevated cosmic-ray ionization rate of $\zeta = 5.0 \times 10^{-15}$\,\rs.
As reference, we take the radiation field found by \citet{spaans:M82} for M82
to model the CO$^+$ abundance found by \citet{fuente:co+}. Their best-fitting
model was an XDR with a density of $n=10^5$\,\ccm\ and an incident X-ray
flux of $F_X=5.1$\,erg\,\rs\,\scm\ ($G_0$=10$^{3.5}$). Using this density
and radiation field, the model predicts the column densities listed in
Table~\ref{t:model}. The chemical structure of the model is shown in
Fig.~\ref{f:model}. 
Models for volume densities far below $n=10^5$\,\ccm\ fail to produce
appreciable \hhhop\ abundances, which makes it unlikely that the origin of the
observed emission is in interclump gas, seen e.g.\ as the $N_H$=10$^{20}$\,\scm\
CO component in M82 \citep{mao:m82}.

Table~\ref{t:model} shows that the \hhhop\ column density observed in M82 may be
produced in either XDR or in the high-$\zeta$ PDR. The \hhhop/\hh\ ratio in M82
is overproduced in the XDR models and underproduced in the standard PDR, but
agrees very well with the high-$\zeta$ model. None of the models reproduce the
observed \hhhop/\hho\ ratio in M82, probably because the observations refer to
very different volumes. We conclude that the M82 data are matched best by the
PDR model with enhanced cosmic-ray flux.
In contrast, the X-ray models do best in reproducing the observations of Arp~220. 
The \hhhop\ column density of 10$^{16}$\,\scm\ is matched by an XDR model with
$N_H$=10$^{23}$\,\scm, scaled up by a factor 10, for either value of $F_X$. 
The \hhhop\ abundance of \pow{1}{-8} is matched by the high-$F_X$ model for low
$N_H$ or the low--$F_X$ model for high $N_H$.
The \hhhop/\hho\ ratio is matched by an XDR of either $F_X$, where low values of
$N_H$ are preferred. Nevertheless, a contribution by the eastern nucleus is
implausible because of its low mass of dense gas and its low infrared brightness.

In the context of galactic nuclei, a PDR model represents a starburst, where the
far-UV is produced by young massive stars and the cosmic rays in supernova remnants.
The pure PDR thus stands for a young starburst and the high-$\zeta$ model for
an evolved one. The XDR models represent the much harder radiation environment
of an AGN, such as the accretion disk of a supermassive black hole.
We find that the observations of M82 are matched by a high-$\zeta$ PDR, i.e.,
an evolved starburst. This conclusion is consistent with earlier models of this
source \citep{schreiber:m82}.
Our \hhhop\ observations do not require X-ray irradiation as in the case of
CO$^+$ \citep{spaans:M82}.

The fact that the XDR models give conflicting requirements on the $N_H$ value
toward Arp~220 may well be due to the unusual geometry of this object.
The emerging picture of the western nucleus of Arp~220 is that there is a 35 pc
optically thick dust disk surrounding what could potentially be an active
nucleus. The dust disk is warm ($T_{\rm D} \approx$ 170 K) and has a peak H$_2$ column
density of $\approx$10$^{25}$\,cm$^{-2}$. Hence X-rays, optical, and even infrared emission
from the nucleus itself will be absorbed \citep{downes:arp220}. Surrounding
this warm dust disk is a cooler ($T_{\rm k} \approx $ 50 K) molecular disk or
torus causing CO 2--1 absorption features towards the hotter nuclear dust disk
in the line of sight. Most of the observed starburst activity appears to be
associated with the rotating disk/torus surrounding the nuclear dust disk.
Future modeling efforts of \hhhop\ in Arp~220 should take this special geometry
into account. However, high-resolution observations are needed to constrain
these models. It is possible for example that the disk causes both \hhhop\
emission and absorption along the line of sight.

\begin{table}[tb]
\caption{Results of PDR / XDR models. } 
\label{t:model}
\begin{tabular}{rrrrrr}
\hline
\hline
\noalign{\smallskip}
$N_H$ &  $N$(\hh)  &   $N$(\hhhop) &  $N$(\hho) &  \hhhop/\hh &  \hhhop/\hho \\
\scm  &  \scm      &   \scm        &  \scm      &             &  \\
\noalign{\smallskip}
\hline
\noalign{\smallskip}
\multicolumn{6}{c}{XDR $F_X$ = 5.1 erg\,\rs\,\scm} \\
\noalign{\smallskip}
\pow{1}{22} & \pow{1.4}{21} & \pow{6.1}{13} & \pow{5.6}{15} & \pow{4.3}{-8} & \pow{1.1}{-2} \\
\pow{3}{22} & \pow{7.9}{21} & \pow{1.9}{14} & \pow{2.2}{16} & \pow{2.5}{-8} & \pow{8.7}{-3} \\
\pow{1}{23} & \pow{3.9}{22} & \pow{3.6}{14} & \pow{5.8}{16} & \pow{9.2}{-9} & \pow{6.1}{-3} \\
\noalign{\smallskip}
\multicolumn{6}{c}{XDR $F_X$ = 16 erg\,\rs\,\scm} \\
\noalign{\smallskip}
\pow{1}{22} & \pow{1.7}{20} & \pow{3.1}{12} & \pow{1.7}{14} & \pow{1.8}{-8} & \pow{1.8}{-2} \\
\pow{3}{22} & \pow{2.3}{21} & \pow{1.5}{14} & \pow{1.2}{16} & \pow{6.6}{-8} & \pow{1.2}{-2} \\
\pow{1}{23} & \pow{2.5}{22} & \pow{8.4}{14} & \pow{8.9}{16} & \pow{3.4}{-8} & \pow{9.4}{-3} \\
\noalign{\smallskip}
\multicolumn{6}{c}{PDR $G_0$=10$^{3.5}$ $\zeta=5.0\times10^{-17}$\,\rs} \\
\noalign{\smallskip}
\pow{1}{22} & \pow{4.6}{21} & \pow{2.3}{11} & \pow{1.6}{14} & \pow{5.0}{-11} & \pow{1.4}{-3} \\
\pow{3}{22} & \pow{1.5}{22} & \pow{1.7}{12} & \pow{1.6}{15} & \pow{1.2}{-10} & \pow{1.1}{-3} \\
\pow{1}{23} & \pow{5.0}{22} & \pow{6.8}{12} & \pow{6.6}{15} & \pow{1.4}{-10} & \pow{1.0}{-3} \\
\noalign{\smallskip}
\multicolumn{6}{c}{PDR $G_0$=10$^{3.5}$ $\zeta=5.0\times10^{-15}$\,\rs} \\
\noalign{\smallskip}
\pow{1}{22} & \pow{4.6}{21} & \pow{9.2}{12} & \pow{2.7}{15} & \pow{2.0}{-9} & \pow{3.4}{-3} \\
\pow{3}{22} & \pow{1.4}{22} & \pow{4.3}{13} & \pow{1.3}{16} & \pow{2.9}{-9} & \pow{3.2}{-3} \\
\pow{1}{23} & \pow{4.9}{22} & \pow{1.6}{14} & \pow{5.1}{16} & \pow{3.2}{-9} & \pow{3.1}{-3} \\
\noalign{\smallskip}
\hline
\end{tabular}
\end{table}

In the near future, the HIFI heterodyne spectrometer onboard ESA's
\textit{Herschel} space observatory will be able to measure \hho\ and \hhhop\
lines around 1~THz and make accurate estimates of the \hho/\hhhop\ ratios in
many sources. These data will allow a much better understanding of the
ionization rates of molecular regions inside and outside our Galaxy.
Future additions to the model include the effects of freeze-out onto and
desorption from dust grains, the effect of metallicity changes, and (especially
for the Arp 220 merger) the time dependence introduced by shocks.

\begin{acknowledgements}
  We thank the staff of the JCMT for their support, in particular Remo Tilanus,
  Jan Wouterloot, and Jim Hoge, and thank Rainer Mauersberger for the quick and
  useful referee report.
\end{acknowledgements}

\vspace{-0.5cm}

\bibliographystyle{aa}
\bibliography{aa8824}

\end{document}